\documentclass[a4paper,11pt]{article}
\pdfoutput=1 

\usepackage{jcappub} 

\usepackage[T1]{fontenc} 

\title{\boldmath Could primordial magnetic fields be generated by Charged Ultra-Light Boson Dark matter?}

\author[a]{Maribel Hern\'andez M\'arquez}
\author[a]{Ana A. Avilez L\'opez}
\author[b]{Tonatiuh Matos Chassin}

\affiliation[a]{Facultad de Ciencias F\'isico-Matem\'aticas, Ciudad Universitaria, Benem\'erita Universidad Aut\'onoma de Puebla, Av. San Claudio SN, Col. San Manuel, Puebla, M\'exico.}
\affiliation[b]{Depto. de F\'isica, Centro de Investigaci\'on y de Estudios Avanzados del IPN, A.P. 14-740, 07000, Ciudad de M\'exico, M\'exico}

\emailAdd{marihm111@gmail.com}
\emailAdd{aavilez@fcfm.buap.mx}
\emailAdd{tmatos@fis.cinvestav.mx}

\abstract{In this work we study the possibility that primordial magnetic fields observed in galaxies could be  produced by a dark matter halo made of charged ultra-light bosons. In our model, we assume that ultra-light bosons arise as excitations of a complex scalar field described by the Klein-Gordon equation with local $U(1)$ symmetry which introduces electromagnetic fields that minimally couple to the complex scalar current. We use classical solutions of the Klein-Gordon-Maxwell system to describe the density profile of dark matter and magnetic fields in the galaxies. We consider two cases assuming spherical and dipolar spatial symmetries respectively. For the particular case of the LSB spherical galaxy F563, we test the sensitivity of the predicted rotation curves in the charged scalar field dark matter (cSFDM) model to variations of the electromagnetic coupling and, by using the Fisher matrix error estimator, we set a constraint over that coupling by requiring that theoretical rotation curves lay inside the $1\sigma$ confidence region of observational data . 
We find that cSFDM haloes are able to generate  magnetic fields of the order of $\mu G$ and reproduce the observed rotation curves of F563-V2 at the same time if the ultra-light boson has a charge lower than $\sim 10^{-13}e$ for the monopole-like density profile and lower than $10^{-14}e$ for the dipole-like one. }


\begin{document}
\maketitle
\flushbottom

\section{Introduction}
\label{sec:intro}
Magnetic fields are present in all the Universe and they are a common feature of various astrophysical objects at different scales. Fields with strengths of $\mu$G are commonly observed in spiral galaxies and the intra-cluster
medium of clusters of galaxies.
Usually magnetic fields in galaxies are observed indirectly as radio waves coming  from stars with frequencies around $10 GHz$. These radio-waves are  produced as synchrotron radiation by high energy electrons traveling across galactic magnetic fields \cite{Kronenberg}.
From decades a host of techniques have been used to detect them and large efforts have been invested to understand their origin and evolution along the cosmic history. A large set of works suggest that they affect galaxy formation and  cluster dynamics.
The origin of large-scale magnetic fields remains as an open problem and a matter of active discussion and research. A host of scenarios and mechanisms have been proposed in order to explain how the seeds of magnetic fields in galaxies were generated at the primitive stages in the cosmic history and their fate in the later epochs of structure formation. Among the best accepted ideas we distinguish the following: i) Primitive magnetic fields were produced in galaxies between recombination and the start of galaxy formation. According to the prevailing picture proposed by  Harrison in the 1970's \cite{Harrison:1970}, swirls in the primordial baryon-photon plasma --right before irrotational curvature perturbations start to dominate-- could be the seed of large scale magnetic field which would have been enhanced at later times via dynamo mechanisms. However, the amplification of this primeval fields up to orders of $\mu$G via this sort of mechanisms last larger times than a galaxy lifetimes in order to be reconciled with observations  ii) Another more modern possibility is that magnetic fields were indeed a primordial seed arising during the pre-recombination era due to a more fundamental mechanism such as inflation \cite{Kronenberg}.
This work proposes an alternative  mechanism to generate magnetic fields in galaxies. The idea is that magnetic fields in galaxies arise due to a very weak electromagnetic coupling between particles of ultra-light scalar dark matter. In this model, haloes of galaxies are macroscopic Bose-Einstein-condensates (BEC) made of hipo-charged bosons with a mass $\sim 10^{-22}eV$ whose density of particles in the Newtonian limit can be described in terms of classical scalar fields.  Various studies have shown that the energy at which the condensation of these objects occurs is $\sim TeV$ at very early age of the universe \cite{Harko:2011,Fukuyama:2009,Haber:1981}. In contrast to the standard cosmological magnetic seed generation mechanisms mentioned above, within this scenario, the appearance of a macroscopic electromagnetic fields is prompt as an inherent fundamental property of the bosonic system.      Ultra-light scalar dark matter is a strong and well studied alternative to the cold dark matter (CDM) paradigm.  It has been shown that these ultra-light bosons are able to form Bose-Einstein Condensates (BEC) at cosmological scales which make up galaxy haloes \cite{Matos-Urena:2001}.
Pioneer systematic studies of this model were firstly carried out by Matos, Guzman and Ure\~na \cite{Matos-Guzman:1998,matos:1999} and rediscovered later on being dubbed with different names such as: fuzzy DM \cite{Hu:2000}, wave DM \cite{schive:2014} or Bose-Einstein condensate DM \cite{BhomerHarko:2007}  (see also \cite{membrado:1989,spergel:1990,sin:1994,lee:1996,Matos-Guzman:1998,arbey:2002,schive:2014} and more recently \cite{Ostriker:2016}) stressed the relevance of this model among the most viable candidates of dark matter.
The cosmological regime of this model was firstly studied in \cite{matos:2000b,Matos-Urena:2001} where it was found that if the ultra-light bosons have a mass $m \sim 10^{-22}eV/c^2$, then the matter power spectrum presents a natural cut-off which suppresses the small-scale structure formation of haloes with masses $M < 10^8 M_\odot$ \cite[see also][]{schive:2014,Urena:2015,Hidalgo:2017}. The same mass of the ultra-light boson has been constrained from several cosmological and astrophysical observations, for example, from the galaxy UV-luminosity function and reionization data \cite{hlozek:2015}, from the high-redshift luminosity function of galaxies \cite{schive:2015}, from Lyman-$\alpha$ observations \cite{Sarkar:2016,Irsic:2017,Armengaud:2017}, taking into account a self-interaction into the SF potential \cite{RindlerDaller:2013}. However, nowadays there is not a full agreement on the exact estimation of ultra-light-boson's mass. However, in this work we use  $m \sim 10^{-22} eV/c^2$ which provides a structure formation description that fairly agrees with observations.
 One important feature of this SFDM model is that it forms structure in the same way as the CDM model at lineal cosmological scales \cite{suarez:2011,magana:2012}. In \cite{Alcubierre:2002a} it was shown that in this model with such mass, the gravitational collapse of the scalar field forms stable objects with masses of the order of a galactic halo.
	In a series of papers it was shown that SFDM forms haloes of galaxies with core density profiles \cite{matos:2007,bernal:2008,robles:2012, BernalGuzman:2006a, BernalGuzman:2006b}. Numerical simulations of formation of galaxies were performed in \cite{medina:2014} and \cite{bray:2015}, where the process of formation of spiral arms of a galaxy was naturally obtained.   Furthermore, the self-interacting scalar field with $\phi^4$  potential was firstly studied in \cite{matos:2011}. 
The SFDM at finite temperature was firstly setup in 
\cite{robles:2013,medina:2015}, and it was also shown that galaxy's satellites are stable around haloes of SFDM \cite{robles:2015}. More features of the SFDM were analyzed further, for example, lensing was studied in \cite{nunez:2010} and other systematic studies of the scalar field dark matter were performed by \cite{arbey:2002} and more recently by \cite{marsh:2014}.
In this model the scalar field at very early stages of the Universe remains almost constant until its wave-length was smaller than the size of the horizon, at such point it underwent a phase transition and started to roll down a new minimum where the scalar behaves as dark energy.  At the condensation temperature 
close to zero, the scalar reaches the new vacuum and it starts oscillating around it and behaves as dust. For ultra-light masses this temperature could be $T_c\sim$TeV \cite{Matos:2017pee}.
Such symmetry breaking process is usually interpreted as a Bose-Einstein condensation transition of a system of bosons  being clumped by gravity. As such, the order parameter of the phase transition corresponds to the occupation number of bosons in the ground state which, at the Newtonian limit, is governed by the Gross-Pitaevskii-Poisson (GPP) system. Solutions of such equations represent the macroscopic wave-function of the whole system which consists in the sum of  wave functions of individual particles which are identical and therefore scales as the number of particles.  We assume that within some range of applicability, the relativistic version of the GPP equation is the Klein-Gordon equation according to \cite{Haber:1981}. Specifically, the classical field solutions of the KG equation describe the order parameter of a system of bosons in the relativistic regime. 
Given that this model provides one of the strongest dark matter candidates together with unresolved question about the origin of magnetic fields at large scales, we believe that the mechanism proposed in this work must be thoroughly studied, being this work a very first step in such task. 
Now let us describe the mathematical framework behind the model. Usually charged bosons can be described mathematically as modes of a canonical complex scalar field with local $U(1)$ symmetry where a vector field mediates an electromagnetic interaction with coupling $q$. Such a system is governed by the Maxwell-Klein-Gordon Lagrangian.
 The $U(1)$ electromagnetic four-vector field  is also responsible for the electromagnetic interactions between baryons. An intriguing question about these charged bosons is whether the theory is safe from massive production of particles via indirect scattering processes. It is essential to make such a safety check on any candidate of dark matter in order to be consistent with observations at large scales that clearly suggest that dark matter is non (or at least weak) interacting. It is not difficult to notice that SFDM is safe of such a catastrophe because even at large energies --below $MeV$ though-- the amplitude of the scattering process involving incoming bosonic particle-antiparticle into an electron-positron pairs mediated by quantum fluctuations of the photon, is suppressed by a \textbf{$1/M_h^2$} factor where  \textbf{$M_h^2$} is the mass of the heavy fermions (confined quarks inside protons and neutrons) making up baryons.The previous can be clearly realized by inspecting the amplitude at tree level in the perturbative expansion of such a process in the simplest theory where heavy fermions are minimally coupled to photons or alternatively using an effective quantum field theory approach. The theory is also safe of a scattering catastrophe via the t-channel through the massless photon exchange thanks to the crossing symmetry of the $\mathcal{M}$ matrix. These decays should happen before the $e^-e^+$ annihilation in order to keep cSFDM cool at the early universe.\\
 One of the main results of this work is that, in order to preserve the predictions at astrophysical scales such as the stellar rotation curves in galaxies, the electromagnetic coupling of SFDM bosons is strongly restricted to very low values.

Now, the main problem solved in this work is the following: if a galactic halo is made by a condensate of charged bosonic modes of the scalar field, it is reasonable to expect that the number density of particles is affected  by the presence of an electromagnetic field in the galaxy. This change in the density should be reflected on rotation curves of stars and gas in the galaxy. The first part of this work is devoted to investigate whether cSFDM can predict rotation curves of a galaxy and at the same time to give magnetic fields with strengths at the microgauss level. Secondly, we  constrain the electromagnetic coupling of bosons by requiring that the theoretical rotation curves to lay within the $1-\sigma$ confidence region of rotation curves of the F563 galaxy. We chose this galaxy due to its morphology and low brightness that make it a good candidate of a "pure dark matter system". This paper is organized as follows, in section II we present the model of dark matter made of charged ultra-light bosons laying in a Newtonian space-time which 
is surrounded by a thermal bath.
In section III we present the mathematical setup describing the system which corresponds to the Maxwell-Klein-Gordon 
system. We carry out an harmonic decomposition either of the density distribution of dark matter and electric field  and we set an azimuthal symmetric ansatz for the average macroscopic total magnetic potential.  In section IV we construct the specific model for the haloes of cSFDM and the set of equations to be solved numerical for our further analysis. Within this section we make a qualitative analysis of our solutions in order to understand the type of solutions of magnetic fields. In section V  we estimate the parameters of the density profile of the pivotal neutral model used to fit the rotation curves of F563. In section VI, by using the Fisher matrix estimator of errors, we derive bounds for the coupling of the scalar and magnetic fields such that the predicted rotation curves are not discriminated at $1\sigma$ level given the data resolution. Finally we summarize and layout our results  and conclusions in section VII.
\section{Dark Matter as a Complex Scalar Field} 
In this section we present the general classical field equations describing our scalar charged dark matter model. By now we don't specify the geometry of space-time, later at the next section we will justify that a Minkowski geometry is a good approximation for the purposes of this work. The charged scalar modes making up the dark matter haloes can be mathematically described by a complex scalar field with local $U(1)$ symmetry which introduces Abelian  
gauge fields which play the role of a mediating particle of the electromagnetic interaction between charged bosonic modes. The corresponding Lagrangian is given by 
\begin{eqnarray}
\label{lagrangiano}
\mathcal{L}=(\nabla_{\mu} \Phi+i A_{\mu}\Phi)(\nabla^{\mu}\Phi^*-i A^{\mu}\Phi^*)-\frac{m^2}{2}\Phi^*\Phi-\frac{1}{4\hat{\mu}_0}F_{\mu \nu }F^{\mu\nu},
\end{eqnarray}
This Lagrangian has units of $\text{distance}^{-4}$, fields in it have  units of distance$^{-1}$ and they have been defined from the canonical ones as follows:  $\Phi\rightarrow \Phi/\hbar c$ and  $ A_\mu \rightarrow q A_\mu/\hbar$. Besides, the parameter $\hat{\mu}_0\equiv 4q^2c\mu_0/\hbar$ is  dimensionless and quantifies the charge of individual bosons given by $q$. 

The kinetic term for the electromagnetic fields in ~\eqref{lagrangiano} involves the electromagnetic tensor defined as:

\begin{equation}
F_{\mu\nu}=\nabla_{\mu}A_{\nu}-\nabla_{\nu}A_{\mu}.
\end{equation}
For now we assume a non-self-interacting field. In the context of SFDM it also encodes the process of condensation of BEC at the very early universe. However,  once dark matter haloes have formed by condensation, an effective mass term suffices to account for the galactic dynamics.
\section{Maxwell-Klein-Gordon System}

In general, the dynamics of a scalar field is  governed by the Klein-Gordon-Maxwell (EKGM) equations in a fixed curved space-time arisen from the 
Lagrangian ~\eqref{lagrangiano} with the potential 

\begin{equation}
\label{kleing}
\left(\nabla^{\mu}+i A^{\mu}\right)\left(\nabla_{\mu}+i A_{\mu}\right)\Phi-m^2\Phi=0.
\end{equation}
\begin{equation}
\label{divj}
\nabla_{\nu}F^{\mu\nu}=-4\hat{\mu}_0 j^{\mu},
\end{equation}
where  the conserved current of charged bosons is defined as 
\begin{equation}
\label{corriente}
j^{\mu}=i\left[\Phi^*(\nabla^{\mu}\Phi+i A^{\mu}\Phi)-\Phi(\nabla^{\mu}\Phi^*-i A^{\mu}\Phi^*\right)],
\end{equation}
and hence satisfies 
\begin{equation}
\nabla_{\mu}j^{\mu}=0.
\end{equation}
Equation (\ref{kleing}) can be expanded to 
\begin{equation}
\label{kleing2}
\nabla^{\mu}\nabla_{\mu}\Phi+i (\nabla^\mu A_\mu)\Phi  + 2iA^\mu\nabla_\mu\Phi- A^\mu A_\mu\Phi
-m^2\Phi=0.
\end{equation}

Notice that electromagnetic and scalar fields are  coupled, even in absence of the gravitational field which for a galaxy is actually very small. Besides, the coupling between scalar and electromagnetic fields is controlled by the $\hat \mu_0$ dimensionless parameter and the strength of the electromagnetic fields (quantified by the initial conditions for $A_\mu$ as we shall see later). Because the scalar field makes up dark matter, therefore it must interact very weakly with all types of fields including with these "dark" photons. Thus it is expected that $\hat \mu_0<<1$. On the other hand, it can be noticed in equation (\ref{kleing2}) that electromagnetic fields could greatly affect the scalar solution used to model galaxy haloes. The goal of this work is to quantify the maximum charge of the boson allowed such that the gravitational strength of dark matter remains unchanged, given the resolution of data of rotation curves used to measure it. 
 Since the multi-state neutral solution provides good fits for these rotation curves, from the beginning we can expect the charge to be small and the electromagnetic fields as well. By such physical argument, it is reasonable to treat the non-linear terms in the KG equation involving electromagnetic fields as perturbations.
 \subsection{Multipolar Decomposition in the Newtonian Limit}
It has been shown that after condensation, in the weak field regime, solutions of the Schr\"{o}dinger-Poisson system are asymptomatically stationary and homogeneous likewise the gravitational potential \cite{GuzmanUrena2006} .
Therefore, in the Newtonian limit we can assume that the space-time metric has spherical symmetry and is given by
\begin{equation}
\label{eq:Metric}
ds^2=-f(r)\,dt^2+\frac{dr^2}{f(r)}+r^2(d\theta^2+\sin^2(\theta)d\varphi^2),
\end{equation}
where $f$ is the gravitational field given by $f=e^{-2U/c^2}\sim 1-2U/c^2$, being $U$ the Newtonian potential.
Because, we are dealing with a non self-interacting field and we assume that the effects of the EM coupling are small as explain previously, the solutions of the KG equations are going to be quite close to those for a neutral scalar except for  perturbations due to the small non-linearities. A neutral scalar solution governed by KG can be written as \begin{equation}\label{eq:PhiDecom}
\Phi=\Phi_0\mathcal{R}(r)Y_N^M(\theta,\varphi)e^{-i\omega t},
\end{equation}
where $Y_N^M(\theta,\varphi)$ are the spherical harmonics and $\Phi_0$ is a real constant with units $\text{distance}^{-1}$.  
Also, we impose an specific form of the EM four-vector which is compatible with the Lorentz gauge condition, given by
\begin{equation}\label{eq:EMPotDecom}
A_{\mu}=(A_0(r),0,0, A_{\varphi}(r,\theta)).
\end{equation}
Now let us to plug the above relations into the Klein-Gordon equation (\ref{kleing}) for a non self-interacting scalar field, its radial component  $\mathcal{R}$ is governed by
the following ordinary differential equation 
\begin{eqnarray}
\frac{1}{r^2}\left(r^2f\mathcal{R}_{,r}\right)_{,r}-\frac{N(N+1)}{r^2}\mathcal{R}+\frac{(\omega-A_0)^2}{f}\mathcal{R}-\frac{A_{\varphi}(A_\varphi+2M)}{r^2\sin\theta^2}\mathcal{R}+k^2\mathcal{R}=0
\end{eqnarray}

 \begin{equation}\label{eq:GeneralRadial}
 \left(r^2\mathcal{R}'\right)'+\Omega(r,\theta)\mathcal{R}=0
 \end{equation}
 \begin{equation}
 \Omega(r,\theta)\equiv\left[k^2-\frac{N(N+1)}{r^2}-2\omega A_0-\frac{2MA_{\varphi}}{r\sin\theta}+A_0^2-\frac{A_\varphi^2}{r^2\sin\theta^2}\right]r^2
   \end{equation}
where we define the wave-number of the scalar $k$ as  
$k^2 \equiv\hat{\omega}^2-\hat{m}^2_{\Phi}$  and $\hat \omega = \omega/c$.
The electromagnetic fields generated by the system of charged bosons are governed by the Maxwell equations given by ~\eqref{divj} which are sourced by $\Phi$ through the current given as \eqref{corriente}. These equations reduce to the following form after plugging ~\eqref{eq:PhiDecom} and ~\eqref{eq:EMPotDecom} in ~\eqref{divj}
\begin{eqnarray}
-\nabla^2 A_0 &=& -4\hat{\mu}_0j_0\\\label{eq:EMspatial}
-\nabla^2 A_k +g^{kk}\partial_k \partial_\phi A_\phi  &=& -4\hat{\mu}_0j_k\\
\nonumber
k=\{r,\theta,\varphi\}&&
\end{eqnarray}
Clearly the temporal and azimuthal components of the current $j_0$ and $j_\phi$ are non-trivial sources of the Maxwell equations since their corresponding $A_\mu$ are non-vanishing. Notice that because $A_r=0$ and the radial component of $\Phi$ is real then $j_r=0$. However, meanwhile $M>1$ the polar component of the four-current will be non-vanishing even if $A_\theta=0$. Since we are studying the simplest case for electromagnetic fields generated by a charged scalar,  (\ref{eq:EMPotDecom}) is setup such that $A_\phi$ does not depend on $\varphi$ and hence $M=0$. Consequently, either $j_\theta$ and the last term on the left of (\ref{eq:EMspatial}) vanish. Then the only non-trivial components of the Maxwell equations are 
\begin{eqnarray}
\nabla^2A_0 &=&  -8 \hat{\mu}_0|\Phi|^2(\hat{\omega}+A_0)\\
\nabla^2A_\varphi &=& -8 \hat{\mu}_0|\Phi|^2A_\varphi
\end{eqnarray}

The previous equations written in spherical coordinates become:
\begin{eqnarray}\label{eq:KGSpherical}
&&\frac{1}{r^2}\frac{\partial}{\partial r}\left(r^2\frac{\partial A_0}{\partial r}\right) + 8\hat{\mu}_0|\Phi|^2(\omega+A_0)=0\\ 
&& \frac{1}{r^2}\frac{\partial}{\partial r}\left(r^2\frac{\partial A_\varphi}{\partial r}\right) + 
\frac{1}{r^2\sin\theta}\frac{\partial}{\partial\theta}\left(\sin\theta\frac{\partial A_\varphi}{\partial\theta}\right)+
8\hat{\mu}_0|\Phi|^2A_\varphi=0
\label{MaxwellSpherical}
\end{eqnarray}
In the next section, solutions of the last system are going to be used to model galactic haloes made of charged bosons which generate magnetic fields of order of $\mu G$. For that purpose we use a specific setup of symmetries and boundary conditions for the system according to the physical situation considered.  
\section{A Model of Dark matter as Charged-Bosons }
Let us recall what the main questions of this work are: how strong may be the electromagnetic interaction between charged bosons such that the model can reproduce the observed rotation curves in a given galaxy? Is it possible to predict observations of kinematics of the visible components of galaxies (rotation curves) and the magnetic fields at the same time?  And consequently, how sensitive are the rotation curves to the electromagnetic interactions between bosons? what bound for the charge of the bosons is allowed such that this conditions are satisfied?  In order to start looking for an answer, we shall construct a simple model that will allow us to investigate the viability of this scenario.
In that direction, we shall use the system of the previous section to model the density profiles of galactic haloes made of charged bosons and the magnetic fields they trigger. The values of the parameters, symmetries and boundary conditions of the system can be reduced according to the phenomenological setup and that is  our goal in this section. 
As mentioned above, because the KGM equations have small non-linear terms and hence the harmonic decomposition of the scalar solution is valid. According to this, here we present two types of solutions of the KGM equations: 1) First, a spherically symmetric complex scalar field coupled to electromagnetic fields decomposed as a magnetic vector potential with constant azimuthal direction and as an homogeneous electric potential. 2) Secondly, a complex scalar field with axial symmetry described by the dipole of the harmonic decomposition and the same setup for EM fields than the first case. We consider this second case in order to account for rotating haloes that could give rise to whirls in the boson gas able to produce an effective non-vanishing magnetic dipole in galaxies. 
\subsection{Angular Decomposition of the EM Potentials by Phenomenological Setup}
An important question regarding this model to be answered beyond this work is: is there a possible mechanism  to form microscopic dipoles within a stationary spherical charged scalar configuration?
 In principle, the simplest setup would be to choose by hand an spherical density configuration and a dipole for the magnetic potential (the first order term in multipole expansion of the magnetic field). However for sake of mathematical consistency, it is expectable that the scalar and EM fields to have common spatial symmetry. 
 On the other hand, an interesting physical scenario that may provide a way to set boundary and initial conditions for the magnetic potential is the following: vortices are shown to exist inside rotating BECS, therefore if the bosons are charged then it comes naturally that they are able to produce magnetic dipoles. This scenario provides a natural way to physically implement magnetic dipoles in a stationary scalar configuration, however an important requirement is to have a rotating halo and therefore, strictly speaking spherical haloes do not allow this mechanism. As we shall show in further sections, bounds on the EM coupling $q$ are independent on whether we use a spherical solution of the first multi-polar moment. Nonetheless, it is worth to take  both cases into consideration in order to verify that using the simplest spherical symmetric haloes approach is a valid approximation when studying gravitational effects and the perturbative effects of possible galactic dark EM fields.  Let us make the following field-redefinition  accordingly to the previous arguments:
\begin{eqnarray}
\centering
A_0&=&\frac{\phi}{r}\\
A_\varphi&=&S(r) Y_{10}(\theta)\label{eq:A}
\end{eqnarray}
after plugging the previous definitions into the field-equations they read:
\begin{eqnarray}\nonumber
&&\left(r^2 \mathcal{R}'\right)' +\left(\frac{N(N+1)}{r^2}+k^2+S^2\cos^2\theta - \frac{\phi^2}{r^2}-2\hat\omega \frac{\phi}{r}\right)\mathcal{R}=0  \\ \nonumber&&\\\nonumber
&&\phi''+8\hat\mu_0\Phi_0^2R^2(\hat\omega r + \phi)=0\\\nonumber&&\\\nonumber
&&(r^2S')+\left(-2+ 8\hat{\mu}_0\Phi_0^2R^2r^2\right)S=0\\
\label{eq:Smagnetic}
\end{eqnarray}
Next we are going to solve this system numerically for the two cases described at the beginning of this subsection. For that purpose we wrote our own code in python using a equivalent version of the system above but using fully dimensionless variables and parameters for better numerical performance. 
\subsection{Solutions of the Maxwell-Klein-Gordon System}
Now let us study qualitatively the behavior of the solutions of (\ref{eq:Smagnetic}) for some regimes of the space of parameters. Afterwards we are going to focus in a range of them suitable for modeling galactic haloes and their magnetic fields. Let us recall once more that the effects of the EM fields are small in order to respect the rotation curves, therefore the neutral model where they are absent is going to play the role of a pivot or fiducial model with respect to which changes due to EM fields are going to be compared.  
\subsubsection{Behavior of the EM Potentials Along the Galactic Plane}
\label{sec:phases}
\begin{figure}[htbp]
 \centering
 \includegraphics[width=12cm]{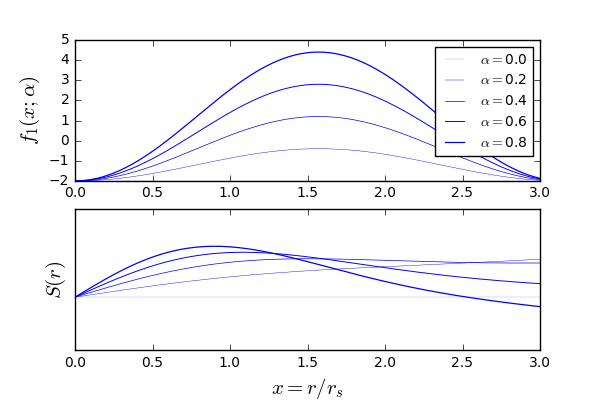}
\caption{$f_1(r;\alpha)$ (top) along the galactic disc for different values of $\alpha$ and their corresponding $S$ solution (bottom). Notice that small values of $\alpha$ give rise to negative $f_1$ which correspond to monotonic solutions while solutions with large $\alpha$ have a more complex oscillatory behavior.} 
\label{fig:Osc2}
\vspace{5mm}
\end{figure}

In this section we aim to study the qualitative behavior of the magnetic potentials generated by the charged bosonic particles for different regimes of parameters. Typically the magnitude of electrostatic potential is subdominant --by a huge gap in orders of magnitude -- in comparison to the magnetic potential, then we are going to ignore it from now on. 
In addition, solutions for the spherical neutral halo have the following form $\mathcal{R}\sim\sin(kr)/r$, therefore by plugging it into (\ref{eq:Smagnetic}) (only by now,later we shall use the exact solution arisen from the coupled system)
we arrive to the following equation for the magnetic potential.
By changing $S\rightarrow S/r$ we would have 
\begin{eqnarray}\nonumber
S''&+&f_1(x;\alpha)S=0\\\nonumber
f_1(x;\alpha)&\equiv&  8\alpha\sin(x)^2-2\\
\alpha&\equiv&\hat{\mu}_0\Phi_0^2
\label{eq:Smagneticf1}
\end{eqnarray}
Where double dots denote second derivative w.r.t. $x=kr=r/r_s$.
 The point where the sign of $f_1$ flips corresponds to a  turnover point from which the solutions start to oscillate after being growing monotonically. 
 Figure (\ref{fig:Osc2}) shows that for small $\alpha$, $f_1(x;\alpha)$ is negative and its corresponding solution for the magnetic potential is monotonical for small radii. In contrast, modes with large $\alpha$ have a more complex evolution at the centre and decay  for larger radii. Let us notice from (\ref{eq:Smagneticf1}) that $\alpha$ corresponds to a simultaneous measure of the EM coupling of the scalar and the central density of the halo, therefore even for small values of the charge, haloes with sufficiently high central density might give rise to large $\alpha$.\\
\subsubsection{Initial Conditions for Magnetic Fields}

Boundary conditions of fields in (\ref{eq:Smagnetic}) should be setup in correspondence to the physical features of the system to be modeled. For example, they should depend upon the charge in order to turn the EM-fields off whenever the bosons are charged. Also, EM fields should decay asymptotically. A simple boundary condition is an effective dipole formed by a charge unit with velocity $\mathbf{v}$ moving around a circle with radius $r_0$. In this case, the magnetic moment would be   
\begin{equation} 
\boldsymbol{\mu} = \frac{q}{2} (\mathbf{r}_0 \times \mathbf{v}) =  \frac{qr_0v}{2}\mathbf{\hat{v}}\times\mathbf{\hat{r}}
\end{equation}
The corresponding magnetic vector potential is given as
\begin{equation}
\mathbf{S}_0 = \frac{\mu_0}{4\pi r_0^2}  \boldsymbol{\mu} \times \mathbf{\hat{r}}
\end{equation}
Assuming that $\mathbf{v}= v \boldsymbol{\hat{\varphi}}$ and $\mathbf{r}_0=r_0 \mathbf{\hat{r}}$ then 
\begin{equation}
\mathbf{S}_0=\frac{\mu_0qv}{8\pi r_0} \boldsymbol{\hat\varphi}.
\end{equation}
Now, in order to have a dimensionless initial condition let us multiply the last equation by $q/k\hbar$ to obtain \begin{equation}
\mathbf{\hat{S}}_0 = \frac{\hat{\mu}_0}{32\pi}\frac{v}{c}\frac{1}{kr_0}\boldsymbol{\hat{\varphi}}=\hat{S}_0\boldsymbol{\hat{\varphi}}
\end{equation}
Our numerical results turn to be not too sensitive to the magnitude of the initial $\hat{S}_0$ however, the previous setting of boundary conditions is important in order to realize how $S$ depends on the electromagnetic coupling $q$. 
On the other hand, initial conditions for the  derivative of $S$ will determine the order of magnitude of the magnetic field at the end, thus initial conditions for the dimensionless variable used here is given by:
\begin{equation}
S_0'= \frac{\,\,e\,r_s^2}{\hbar}(\beta\mu G)
\end{equation}
where $\beta$ is a fitting parameter which determines the order of magnitude of the magnetic field at the galactic centre and its decaying rate along the disc. In this work we fix it to $\beta=10^{-30}$ in order to produce magnetic fields running within $1-10\,\mu G$. 
\subsection{Rotation Curves and Magnetic Fields from Complex SFDM Solutions}
In this subsection we show the method to compute astrophysical  observables involved in our study in terms of the solutions of the KGM system presented in previous sections. 
In SFDM models, classical scalar solutions of the Gross-Pitaevskii-Poisson (GPP) system are commonly used to model the density distributions of dark matter in galaxies. In the Newtonian limit where such system is valid, it is true that the corresponding density of dark matter can be computed from the scalar solution as \begin{equation}\label{solution}
 \frac{4\pi\,G\rho}{c^2}=\Phi\Phi^*=\frac{\bar\rho}{k^2} \mathcal{R}^2,
\end{equation}
where $\Phi^*$ corresponds to the complex conjugate of the scalar field $\Phi$. For convenience, we set the relation above between the density factor $4\pi G/c^2$ and the scalar field in order to have units of $1/distance^2$. Equivalently, in units where $4\pi G=1$ and $c=1$ , $\rho=|\Phi|^2$ barely.We also have defined the central value density as $\bar{\rho}$.
 Boundary conditions on the radial scalar solution need to be established at this point. The monopole scalar solution for neutral bosons --that is when EM are absent--  which has been typically used to fit rotation curves within this model \cite{robles:2013} has the form $R\sim \sin(kr)/r$, where $k$ is an integration constant to be fixed by boundary conditions. Given that such function is periodic, there is a radius $r'$ where the density distribution vanishes $\rho(r')=0$, this happens if $\{k_jr'=j\pi\}_{j=1,2,...}$. In this way it is realized  the existence of excited states making up the whole solution as a specific superposition of them to be fixed by observations owing that equation (\ref{eq:GeneralRadial}) (with $A_\varphi=A_0=0$) is lineal allows us to form haloes by superposition of excited states as follows
 \begin{equation}
 \phi_{neutral}= \sum_i \phi_{i0} \frac{\sin(k_ir)}{k_ir}. 
 \end{equation}
 It is worth to point out that because the previous form of the solution has a physical interpretation: in the ideal case, solutions of the GPP or SP systems are supposed to describe
perfect BECs with temperature quite close to zero and the system of bosons would condensate to a single macroscopic wave function and all the bosonic excitations would lay 
in the ground state described by the single ground state solution with no spatial nodes \cite{robles:2013}. However, moving one step beyond such idealization, because the system is surrounded by a thermal bath therefore it is expected that excited states arise. 
By virtue of the smallness of EM fields involved in the KGM system above, we assume that the scalar solutions are equal to those of the neutral scalar plus small perturbations and therefore we fix the boundary conditions in the same way. 
\begin{sloppypar}
  In order to compute mass contribution from the density distribution of each multi-state, we should integrate its corresponding density distribution over all space at a fixed time as follows 
\end{sloppypar}
\begin{equation}
M_i(r)= \frac{\bar\rho_i}{ k_i^2}\int_0^r{\mathcal{R}_i(r')^2}r'^2dr'.
\end{equation}
The total mass of the halo corresponds to the sum of all the contributions as follows
\begin{equation}
M_T(r)=\sum_{i}M_i(r)
\end{equation}
\begin{sloppypar}
In the Newtonian regime, the outer movement of stars in a galaxy is mainly governed by geometry of the potential well produced by the halo, though there are sub-dominant effects coming from stellar and gas dynamics deforming the dominant well potential, for the purposes of this work we let them out of consideration. Therefore, 
by using the virial relations we can infer that the magnitude of the rotation velocity of stars can be computed as follows
\end{sloppypar}
\begin{equation}
\label{ecv}
V^2=\frac{M_T(r)}{2\pi r}
\end{equation}
Let us recall here that we are not using international units until  now, however, later on after numerically computing these normalized rotation curves and magnetic fields we shall transform our quantities to those units in order to compare with observations. 
We compute the magnetic field by taking the curl of ~\eqref{eq:A} and we get:
\begin{equation}
\label{B}
\mathbf{B}=\sum\nabla\times(A^i_{\varphi}\boldsymbol{\hat{\varphi}})=B_r\mathbf{\hat{r}}+B_{\theta}\boldsymbol{\hat{\theta}},
\end{equation}
whose components are given as
\begin{eqnarray}
\label{Br}
B_r&=&\sum_i \cos(2\theta)\frac{S_i}{r\sin\theta}\\
\label{Btheta}
B_{\theta}&=&-\sum_i{\left(\frac{S_i}{r}+S'_i\right)\cos\theta}.
\end{eqnarray}
Where we are adding up contributions to the  magnetic fields arisen from all multi-states required to fit the rotation curves.
Notice that either $B_r$ and hence $|\mathbf{B}|$ diverge at $\theta=n \pi$ with $n$ being an integer and at the galactic centre. In what follows we shall compute observables in the galactic disc plane at $\theta=\pi/2$.
\section{Characterizing the Density Profile of the Pivot Dark Matter Halo}
\begin{figure}[htbp]
\centering
 \includegraphics[width=11cm]{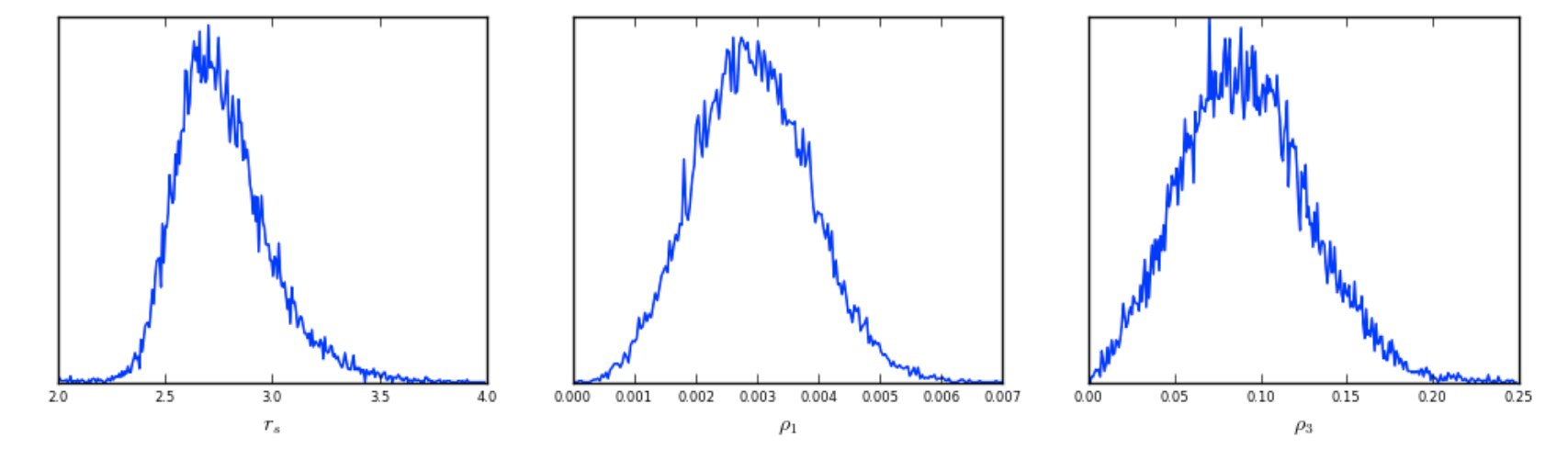}
\caption{Marginalized posterior distributions for the parameters characterizing the halo derived from the monopole of the neutral scalar field. }\label{fig:Post1}
\vspace{5mm}
\end{figure}

\begin{figure}[htbp]
 \centering
 \includegraphics[width=11cm]{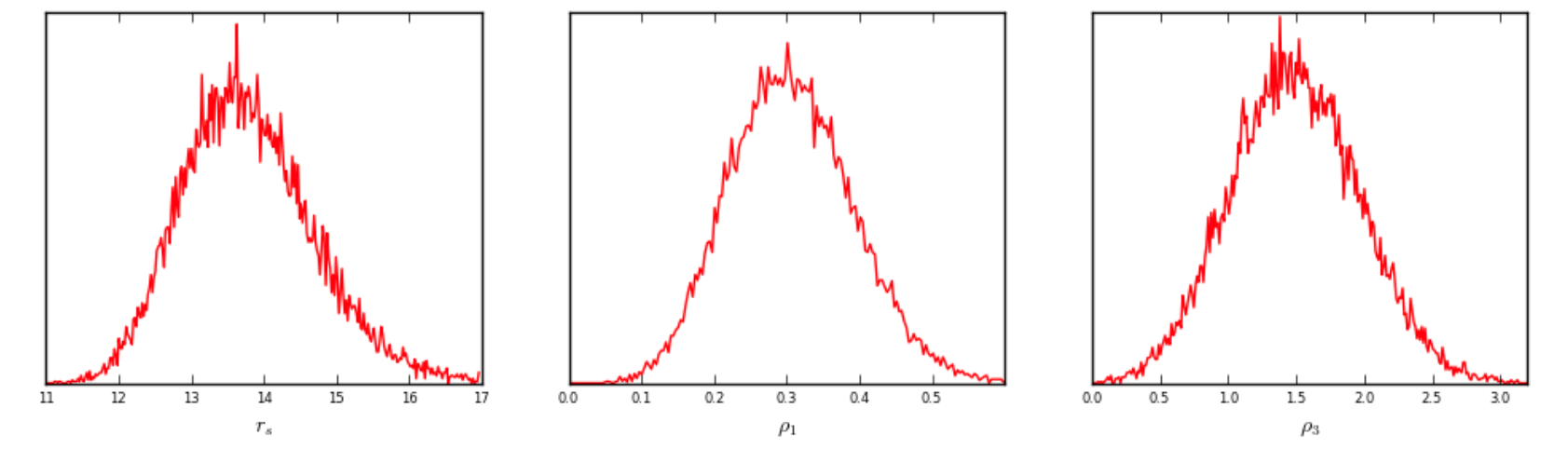}
\caption{Marginalized posterior distributions for the parameters characterizing the halo derived from the dipole of the neutral scalar field.}\label{fig:Post2} \vspace{5mm}
\end{figure}

In this section we determine the density profiles of the scalar field configurations with different spatial symmetries, that is, the monopole and the dipole (recall that the first one is intended to model a rotating halo and the second is the simplest approximation), we estimate the parameters of an halo whose density profile is describe up to three multi-states, with that purpose we fitted the rotation curves of a low-surface-bright galaxy with spheroidal morphology, which is the ideal type of galaxy to test gravitational effects of dark matter onto the stellar kinematics since effects of baryons are presumably sub-dominant and therefore the minimum disc hypothesis is valid. By using measurements of the rotation curves of galaxy  F563 made by \cite{McGaugh:2016leg}, we sampled the space of parameters of the mentioned fiducial models by using our own Monte-Carlo-Markov-Chain code via the Hastings-Metropolis algorithm. The set parameters considered in our analysis are given by: the characteristic size of the halo $r'$ corresponding to the size of the first spatial oscillation of the ground state, the central density of each multi-state $\bar{\rho}_i\qquad i=1,2,3.$. As mentioned above, this model will play the role of a pivot or reference model and the following estimates will lead to an order-zero rotation curves against which rotation curves arisen from charged haloes are going to be compared in order to set bounds on the charge of the bosons in the next section. 
Resulting marginalized posteriors of the parameters are shown in Figures (\ref{fig:Post1}) and  (\ref{fig:Post2}). Distributions for the second multi-state are omitted since it turns out to be needless to fit the rotations curves. Best-fitting parameters for each instance are summarized in table (\ref{tab:params0}).
An interesting outcome from these results is that haloes produced by spherical solutions are smaller than those arisen from dipole solutions. Since multi-state solutions are denser for the dipole scalar than the monopole ones. These fact should be taken into account when one is aiming to carefully describe the shape and dynamics of SFDM haloes, since arbitrarily chosen spatial symmetries of the scalar configuration  could lead to different results. As we shall see in the next section, the spatial symmetry of the scalar importantly affects the upper bounds to the charge $q$, in the case of the monopole larger values of $q$ are allowed than for the dipole solution. An explanation for this in physical terms is that the dipole solution holds orbital angular momentum which has a contribution to the magnetic field and hence, in order to reproduce the rotation curves the charge must be cut down.    
\begin{table}[h!]
\centering
\begin{tabular}{|c|c|c|}
\hline
Multipole & Multistates &Central densities $(M_{\odot}/pc^3)$\\
\hline
&&\\
Dipole&1,3 & $\bar\rho_1=0.29,\,\,\bar\rho_3=1.52$\\ 
&&$r_s=13.67\,kpc$\\
&& \\
Monopole&1,3 &$\bar\rho_1=0.004,\,\,\bar\rho_3=0.078$\\ 
&&$r_s=2.9\,kpc$\\
\hline
\end{tabular}
\caption{Estimates for the parameters of the multi-states yielding to the density profiles of dark matter with dipole and monopole spatial distributions respectively.}
\label{tab:params0}
\end{table}
\section{Bounds of the Boson Charge From F563-V2 Rotation Curves Data}
\subsection{The Method}
 Our main goal now is to estimate the order of magnitude of the charge of the boson $q$ such that predictions of the rotation curves remain inside the error bars of rotation curves ($1\sigma$ of confidence level) and at the same time the resulting magnetic fields lay in a range of micro-Gauss. In other words, the question here is: what is the maximum allowed value of $q$, such that the corresponding theoretical curves lay inside the $1-\sigma$ resolution of rotation curves data?  
Since stellar rotation curves in galaxies provide a measurement of the Newtonian gravitational potential well produced by DM, the cutoff $q$ provides limit from which the gravitational interaction remains ''flawless'' owing to EM interaction between bosons, as far as data is able to resolve.  In other words it quantifies the extent in which the strength of the bosonic electromagnetic coupling affects the number density, and consequently the rotation curves of stars and gas traveling along the potential well detectable so far for the instance of F563.

\begin{figure}[htbp]
\centering
 \includegraphics[width=7.5cm]{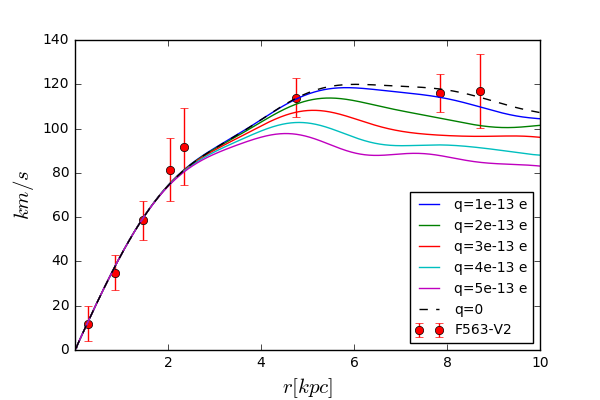}
  \includegraphics[width=7.5cm]{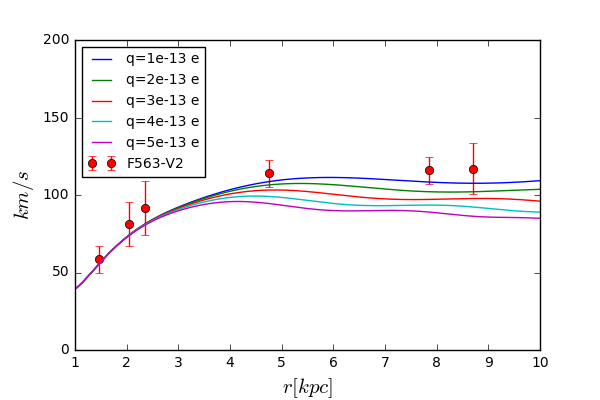}
\caption{Rotation curves derived from the coupled system of scalar and EM field equations. Left panel show rotation curves derived from the monopole solution while the right panel corresponds to the dipolar scalar configuration. Parameters characterizing the density profile of dark matter are those of the pivot neutral model and the value of the charge as indicated in plot labels. }\label{fig:rot}
\vspace{5mm}
\end{figure}

Figure (\ref{fig:rot}) shows theoretical rotation curves arisen either from monopolar and dipolar scalar solutions corresponding to F563 for different values of $q$ laying well inside the 1$\sigma$ confidence region of data. 

In addition, the magnitude of the total magnetic fields produced by the system of bosons run from $1-10\mu$G at the galactic plane. Although these magnetic fields decay within radii of few parsecs in the galactic disc, notice from the first equation in system (\ref{eq:Smagnetic}) that the coupling to $S$ vanishes at $\theta=\pi/2$ and hence the module of the magnetic fields is expected to take its minimal values at the disc. We also verified this numerically. Therefore larger values of the magnetic fields would be measured in regions off the galactic plane for larger regions.
We estimated the bounds on the charge of the bosons $q$ by using the Fisher matrix technique \cite{NumericalRecipes:2007,Heavens:2014} which enables one to estimate errors in terms of variations of theoretical predictions with respect to the theoretical parameters and the observational errors. Consider the theoretical predictions of a set of observables ${\Omega}_k$ within a parametrized set of models by a set of  parameters $P_i$. An estimator of the covariance matrix of such quantity around the fiducial  model with parameters $P_{0i}$ is given by the inverse of the Fisher matrix whose components are computed as \begin{equation}
\mathcal{F}_{ij}= \sum_{a}\frac{1}{\sigma_a^2}\frac{\partial\Omega_a }{\partial P_i} \frac{\partial\Omega_a}{\partial P_j}
\end{equation}
where $\sigma_a$ is the observational Gaussian error associated to $\Omega_a$. In our case, the observables are different data points of the rotation curves and the fiducial model is given by those of the pivot model and $q=0$. Thus, according to this method, an estimation for the error of $q$ would be given by \begin{equation}
\sigma_q=\left[\sum_i \frac{1}{\sigma_{V_i}^2}\frac{\Delta V(r_i)}{\Delta q}\frac{\Delta V(r_i)}{\Delta q}\right]^{-1/2}
\end{equation}
where $\frac{\Delta V(r_i)}{\Delta q}$ are computed using the coupling values $q$ which push the rotation curves to the $1\sigma$ limit and $\sigma_{V_i}$ is the corresponding error bar of the i-th point. The resulting estimates of $q$ corresponding to different sorts of scalar configurations are summarized in table (\ref{eg}).
\begin{sloppypar}

Notice that the theoretical rotation curves decrease as the charge increases. A qualitative explanation for this to happen is the following: as the charge of bosons increases, they repel from each other producing a decrease in the density described by $|\Phi|^2$. The rotation curves of low-surface-bright galaxies are signatures of the depth of the gravitational potential well of dark matter mainly, therefore, if the density of dark matter decreases it is expectable that the rotation curves do as well.  
\end{sloppypar}
\begin{table}[h!]
\centering
\begin{tabular}{|c|c|}
\hline
Multipole&Bound\\
\hline
 Monopole  & $\qquad1.45\times10^{-13}\,e\qquad$\\
  Dipole  & $\qquad 3.28\times10^{-14}\,e\qquad$\\
\hline
\end{tabular}
\caption{Bounds of the charge of the bosons required to the rotation curves to lay inside the 1$\sigma$ confidence region of  F563 data. }
\label{eg}
\end{table}

\section{ Conclusions}

In this work we consider the hypothesis that dark matter haloes are BEC made of charged bosons with mass around $10^{-22}eV$ modeled as classical scalar field configurations which are able to explain not only the rotation curves and the shape of the galaxies but furthermore galactic magnetic fields arisen from them. At first approximation, we show that when the classical coupling constant between the scalar and  electromagnetic fields takes tiny values lower than $\sim 10^{-13}$ times the charge of the electron, by holding  specific boundary conditions, the scalar configuration produces magnetic fields at the galactic centre with magnitude of $\sim\mu$G like those typically observed in  galaxies. In general, the scalar and electromagnetic field equations are coupled, as a consequence their solutions used here to model the density profiles of charged boson haloes might able to alter the rotation curves of galaxies in comparison to those arisen from  neutral configurations. However for bosons holding  charge values below the bound derived here,  both predictions of the rotation curves are indistinguishable by data since they remain inside the 1$\sigma$ confidence region. Of course, this result may change for other galaxies and more precise data in the future, nevertheless we probe the extent of sensitivity of data to the charge parameter and furthermore we show that it is possible that cSFDM is able to produce primordial magnetic fields $\sim \mu $G. Also, further research will be needed in order to determine whether other phenomena comes out from this sort of dark matter and other constraints for this coupling are to be imposed by testing predictions of other cosmological and astrophysical observables.
In order to model such dark matter haloes we used the classical system of KGM coupled equations with local $U(1)$ symmetry providing a minimal coupling between the EM fields and the scalar current. We carried out an harmonic decomposition of our variables given the symmetries of our physical system and that the system of equations hold small non-linear terms involving weak EM fields. We study the simplest spatial distributions of density profiles of the scalar field, that is, spherical and dipole configurations. Dipole-like solutions are physically interesting in order to take into account rotating scalar configurations that are able to produce magnetic fields as a non vanishing macroscopic dipole as a boundary condition arising due to vorticity phenomena in the halo. According to a qualitative analysis of the equation governing the magnetic potential, cuspy scalar fields at the centre and large magnitudes of the boson charge can give rise to complex behavior of the magnetic potential while small values of these quantities turn into monotonic solutions.
We solved numerically the KGM system by setting, as boundary condition, the central magnetic potential as that of a magnetic dipole formed by spinning bosons around a fixed circle, by now we fitted the size of such circle in order to obtain magnetic field around $\mu$G. Further research about vorticity phenomena within cSFDM model will provide information in order to better set these boundary conditions in the future. Afterwards we took, as a case of study, the F563 galaxy with spherical morphology and low-superficial-brightness which are desirable features of a dark-matter dominated system. We took density profiles corresponding to neutral scalar solution as pivot or fiducial models in order to fit the rotation curves F563 and compare them to those arising from cSFDM models in order to derive a bound $q$ for the charge of the bosons such that the rotation curves start to be indistinguishable provided the data. We estimate this error by using the Fisher matrix method. Spherical and dipolar scalar configurations lead to different rotation curves and the bound of $q$ for the former is larger than the latter for one order of magnitude. 

The main conclusion of this work is therefore, that in the case of galaxy F563, a spherical(dipolar) dark matter halo made of cSFDM bosons with charge $\sim 10^{-13} e$($\sim 10^{-14}e$) is able to generate $\mu$G magnetic fields and at the same time to predict the observed rotation curves.

\acknowledgments

This work was partially supported by CONACyT M\'exico under grants CB-2011 No. 166212, CB-2014-01 No. 240512, Project
No. 269652 and Fronteras Project 281;
Xiuhcoatl and Abacus clusters at Cinvestav, IPN; I0101/131/07 C-234/07 of the Instituto
Avanzado de Cosmolog\'ia (IAC) collaboration (http://www.iac.edu.mx/).  M.H. acknowledge financial support from CONACyT doctoral
fellowship. A.A. acknowledge financial support from CONACyT postdoctoral fellowship. Works of T.M. are partially supported by Conacyt through the 
Fondo Sectorial de Investigaci\'on para la Educaci\'on, grant CB-2014-1, No. 240512

\bibliographystyle{JHEP}
\bibliography{MagneticGalaxiesSFDM}
\end{document}